\documentclass[12pt, article]{revtex4}                                                                                                                                                                                                                                                                                     
\usepackage[dvips]{graphicx,color}
\usepackage{multirow,newlfont,enumerate} 
\usepackage{epsfig,epstopdf,ifpdf, latexsym, slashed}
\usepackage{color, graphicx, setspace, xspace} 
\usepackage{amsmath,amssymb,amsfonts,amsthm,amsxtra,subfigure}
\usepackage{todonotes} 
\usepackage{tensor} 
\usepackage{mathtools, array} 
\theoremstyle{plain} 
\newtheorem{thm}{Theorem} 
 
\theoremstyle{definition} 
\newtheorem{defn}{Definition} 
\newcommand{\beq}{\begin{equation}}
\newcommand{\eeq}{\end{equation}}
\newcommand{\beqr}{\begin{eqnarray}}
\newcommand{\eeqr}{\end{eqnarray}}

\def\lsim{\raise0.3ex\hbox{$\;<$\kern-0.75em\raise-1.1ex\hbox{$\sim\;$}}}
\def\gsim{\raise0.3ex\hbox{$\;>$\kern-0.75em\raise-1.1ex\hbox{$\sim\;$}}}

\def\cc{\cfrac{1}{2} } 
\begin{document}
\title {Meromorphic Behavior  of Time Dependent  Schroedinger Equation } 
\author{R. Dutta  \footnote{\it dutta.22$@$osu.edu, Corresponding author}  } 
\author{ A. Stan \footnote{\it stan.7$@$osu.edu}} 
\affiliation{ Department of Mathematics, The Ohio State University  }  
 
\maketitle 

\section{Abstract}

We try to obtain meromorphic solution of Time dependent  Schr$\ddot{o}$edinger equation which partially satisfies Painleve Integrable property.  
Our  study and analysis exhibits meromorphic behavior of  classical particle trajectory. 
 In other words, particle  is confined  in punctured complex domain  in singular fundamental domain. . 
We have explicitly  developed  solution from Jacobi Elliptic function and pole expansion approach in which solution remains meromorphic . 
 Branch point analysis also shows solution branches out near such singular point. 
 Meromorphic behavior is significant for a classical particle within quantum limit. 
 
\section{Introduction} 
Time Dependent Schr$\ddot{o}$edinger Equation (TDSE) is a linear partial differential equation that governs wave function of a quantum mechanical system. 
It is a key result of quantum mechanics and  Hamiltonian of the system.  Being linear PDE,  TDSE  represents integrable PDE. 
Conceptually, this is the quantum counterpart of 
Newtons second law in classical mechanics. Given a set of known initial conditions, 
Newtons law makes a mathematical prediction as to what path a given physical system will take over time 
( using variational method ) . Schr$\ddot{o}$edinger equation yields evolution over time of wave 
function associated with a particle in quantum limit[ ruled by uncertainty principle]. 
A differential equation is always said to be integrable if it is solvable ( for a sufficiently large class of initial data ) via an associated ( single valued )
 linear problem. Non linear  ODE known as autonomous Ermakov - Pinney (EP)  equation is   derived from TDSE representing particle motion does not follow integrable property. 
 Following  ARS conjecture,  which states that {\it any ODE which arises as a reduction of an integrable PDE possesses the Painleve property, 
 possibly after a transformation of variables }.  
 It is to note that during reduction process, TDSE may loose  integrability \cite{kruskal}. dynamics.   
 TDSE is a integrable system  in nature following conservation of hamiltonian of the system. 
 Non linear EP equation is very well known in quantum mechanics as it is derived from TDSE  and 
  physicists such as  [ \cite{schuch} \cite{lewis} ]  tried to establish  connection with non linear classical mechanics. 
 They argued that linear behavior of quantum mechanics is [ where superposition principle is applicable] in disguise 
  non linear in nature.   Joos  et al   \cite{Joos} studied separately  decoherence and appearance of classical mechanics in quantum theory. 
  An invariant dynamical quantity following conserved Hamiltonian  was obtained in evolution of such particle path under harmonic oscillator potential problem.  
  The wave packet solution of the TDSE  for Harmonic oscillator with time dependent frequency yields a coupled pair of equations of motion 
  corresponding to  mean value and uncertainty of position of the particle under wave packet which corresponds to an exact dynamical invariant. 
 First equation describes mean value of momentum in the direction of wave propagation and linear ODE.
  The second ODE describing particle position uncertainty within wave packet ( complex quantity) is non linear ODE  and is known as EP  equation. 
   This ODE is governed  by oscillator frequency only.  (EP is a nonlinear ODE due to third term and is a special case of $P_{II}  $ 	(2nd Painleve equation ) which does not satisfy PP. 		
 The authors [ \cite{schuch} , \cite{lewis} ] obtained solution of  EP equation   in terms of  linear solutions following Pinney calculation. 
Their solution does not exhibit  behavior near singularity. 
 If any autonomous  ODE admits any kind of singularity in terms of unbound  behavior or singularity that arise from the coefficients. 
The singularity can be fixed/ movable type or as branch point in the ODE or  in the solution.  The solution 
is called Meromorphic if the solution function satisfies
\begin{defn} 
 A function $ x \rightarrow u(x) $ from $ \mathbb{C} $ to $ \overline{ \mathbb{C}} $ is called meromorphic on $ \mathbb{C} $ if its only singularities 
 in $ \mathbb{C} $ are poles. 
\end{defn} 
Meromorphic structure of solution for a complex variable is very crucial to study behavior of solution near such singularity. Structure of singularity is also very significant to develop 
method of solution. 
Mathematicians such as  Hoyer , Conte [  \cite{hoyer}  , \cite{conte} ]  studied to develop various types of  general solution in such cases.  
This meromorphic assumption consists of checking 
 existence of Laurent series and its ability to represent general solution i.e to contain enough arbitrary parameters. Since Laurent series is only defined 
within its annulus of convergence , therefore study is only local . 
  From mathematical analysis point of view,  integrability ( solvability via an associated single valued linear problem) 
  of a differential equation is closely associated to singular 
 structure of its solution. In particular,  there is strong evidence that all integrable equations have Painleve property . 
 This means all possible  solutions of the ODE are single valued around all movable singularities. 
 Structure all movable singularities  are defined as pole in which ODE has analytic structure of solution.
  Movable here means that the singularity position varies as a function of initial value and can be designated by integration constant. A differential equation
  is said to have Painleve property  if all solutions are single valued around all movable singularities.   
 There exist six  classes of  nonlinear  ODEs which obey  Painleve integrable property . These ODE s represent  prototypical examples of integrable ODEs. 
 All six classes  posses  characteristic singularity . There is strong evidence  \cite{ziglin}, \cite{yoshida} that the integrability of a non linear system is intimately related to the singularity structure .
 There exist two privileged subsets of single valued functions , the elliptic functions and meromorphic functions. We will adopt only method to construct meromorphic functions in this work. 
 
 It is to note that Painleve property excludes  branching as point of singularity 
and indicates as  pointer to integrability of ODE.  One can identify pole structure existing in the solution of the differential equation of various order. 
The complex singularity structure of solutions was first used by Kovalevski  \cite{kovalevski} . 
Demina et al \cite{demina} developed algorithm to obtain  explicit expression for various classes of meromorphic solution of autonomous non linear differential equations.
They developed algorithm which can be applied to construct exact meromorphic solutions in explicit form for a wide class of non linear ordinary differential equations. 
In the framework of TDSE,  there exists integrable model Hamiltonian for which  analytical solution to exist in form of two independent linear solutions. Examples related to Harmonic oscillator (HO) 
systems are reviewed by  \cite{kleber} .  The response of a harmonic oscillator to a time dependent external force is the prototype for all time dependent problems in quantum mechanics. 
The equation generally represents equation for wave function associated with the particle ( linear type ) and can be solved by time evolution of power series solution of steady state given by
\begin{gather} 
\psi( x, t) = e^{ - \cfrac{ i t H } {\hbar} } \psi( x , 0) = \sum_n \psi_n ( x) e^{ - \cfrac{ i t H}{\hbar} }
\end{gather} 
where $ \psi_n$ is the nth steady state solution obtained in terms of Hermite polynomial and H is the Hamiltonian of the quantum system. 
If one chooses gaussian wave packet representing the particle in quantum system , then one can obtain particle trajectory equation within wave packet. 
 Original formulation of TDSE 
 is based on classical Hamilton Jacobi mechanics
where the equation  is linear and has features of wave equation(  complex quantity) with superposition principle. 
Derived from   TDSE  [ which is represented by Gaussian wave packet] ,
  ODE  representing particle motion in the direction of wave packet propagation  is linear  and solution is represented by Fourier function.  
   This equation gives information about position of the particle in the direction of wave propagation  at any time and so gives linear momentum information. 
 The other equation representing particle position within wave packet [ related to  position uncertainty of the particle ] is second order non linear ODE,   known as EP  equation. 
 Once solved, it can estimate quantum uncertainties of position, momentum and their correlation directly by calculating related mean values in terms of gaussian wave packet solution. 
 Detail calculations are shown in next section.  
 General Ermakov-Pinney(EP) equation  is of the form 
 \begin{gather}
 \ddot{x}(t) + \omega^2 (t) (t) = \cfrac{1}{ x^3(t)} 
 \end{gather} 
  The solution was given  by Pinney as  \cite{andri} 
  \begin{gather} 
  x(t) = \sqrt{ A u^2 + 2 B u v + C v^2 } 
  \end{gather} 
where u(t) and v(t) are any two linearly independent solutions with constants A,B,C are related according to
\begin{equation} 
B^2 - A C = \cfrac{1}{ W^2} 
\end{equation} 
W,  being wronskian of two independent solutions. 
The equation is very significant in mathematics as it possess algebra of $ sl ( 3, \mathbb{R} ) $ , isomorphic to  
 non compact algebra $ s0  (2,1) $ which geometrically represents rotation on the surface of an hyperboloid of one sheet \cite{leach} . 
 Solution (2) of equation (1)  can be interpreted as first integral of third order equation 
 \begin{gather} 
 \dddot{x} + 4 \omega^2 \dot{x} + 4 \omega \dot{\omega} x = 0 
 \end{gather} 
 which is the normal form of a third order equation of maximal symmetry
Interestingly, all linear and linearizable systems of ordinary differential equations of maximal symmetry posses this algebra such as heat equation, Schr$\ddot{o}$edinger equation.  
Cruz  et al  \cite{cruz}  gave existence of a  dynamical invariant  for this system from conservative Hamiltonian  as 
\begin{gather} 
  I = \cfrac{1}{2} \left[ {( \dot{\eta} \alpha - \eta \dot{\alpha} ) }^2 + {\left(\cfrac{ \eta}{\alpha}\right) }^2 \right]  = constant 
\end{gather} 
 where $\eta(t) $ is the particle position of the particle along the direction of wave propagation and $ \alpha(t) $
  represents particle location within the gaussian wave packet [ 
 position uncertainty of the particle within wave packet at any time]. 
  \cite{cruz} obtained  solution for  $ \alpha(t) $ in terms of two linear operators as   
 \begin{gather} 
 \alpha(t) = \sqrt{ \left( \dot{\alpha_0}^2 + \cfrac{1} {{\alpha_0}^2 } \right) {\eta_1}^2 (t) + {\alpha_0}^2 {\eta_2}^2 (t) \pm 2 \dot{\alpha_0} {\alpha_0} {\eta_1}(t)  {\eta_2 }(t) } 
 \end{gather} 
 with two solutions  $ \eta_1(t) $ and $ \eta_2(t) $ from linear part of the system .  Here  $ \alpha_0 $ is the initial condition. That means  solution of  $ \alpha (t) $ depends on linear solution. 
 In this approach, presence of essential singularity in the ODE is ignored in order to construct solution.  
 The behavior of solution near singularity is  very essential  from mathematical understanding  point of view as well as physical interpretation. Therefore,  ODE   requires mathematical attention to construct 
 solution method. 
 In this paper, we consider EP  equation for particle position within wave packet [ related to position uncertainty estimate ]
 as fundamental ODE to represent particle position uncertainty within WP in TDSE. We try to address unbounded behavior of ODE as fundamental ODE problem and 
 construct local behavior of solution  in  particle  trajectory manifold. 
 EP  equation represents second order autonomous ODE which admits  movable singularity  due to presence of third term. 
 Conte  \cite{conte1} identified few class of non linear ODE s  ( under some mild conditions)  in which solution came as transcendental function.
 In this paper, we focus to construct systematic method to obtain analytic solution near such singularity.

 \section{Problem and Analysis}
 TDSE equation under  harmonic oscillator potential in quantum mechanics is given by 
 \begin{gather} 
i \hbar \cfrac{\partial}{\partial t} \psi_{WP}   (x, t) = - \cfrac{\hbar^2}{2m} \cfrac{\partial^2}{\partial x^2}  \psi_{WP} (x,t) + \cfrac{m}{2} \omega^2 x^2 \psi_{WP} (x,t) 
\end{gather} 
where  $\psi_{WP} $  represents   Gaussian wave packet (WP)   associated with particle moving along -x direction where width of wave packet is in -y plane. 
\begin{equation} 
\psi_{WP}  (x, t)  =N(t) \exp \lbrace  i  \left[  y(t) \tilde{x}^2 + \cfrac{ < p> } { \hbar} \tilde{x} + K(t) \right]  \rbrace
\end{equation} 
where y(t) is complex 
\[ y(t) = y_R(t) + \i  y_I (t) \] 
\[ \tilde{x} = x  - < x>  = x - \eta(t) \] 
with Hamiltonian of the system is  determined by momentum and position of the particle
\begin{gather}
H = \cfrac{ \tilde{p} ^2}{2 m} + \cfrac{ 1}{2} {\omega}^2  {\tilde{x} }^2 
\end{gather} 
Here  classical trajectory of the particle is defined 
\begin{gather} 
 \eta(t) = < x >  =  {\int_{\infty}}^{\infty}  {\psi}^{\star}  x  \psi  dx  
\end{gather} 
with classical momentum 
\[  < p > = m \dot{\eta} \] 
 $ \hbar $ being Planck constant.  Imaginary number i  in equation   appears  as a result of uncertainty property.  N(t) and K(t) 
 being time dependent constants that are not relevant for following analysis. 
 Inserting WP (9 - 10 ) into equation (8), we obtain the equation of motion for $  \eta(t) $ and y(t). 
 The equation for the WP maximum , located at  $ x = \eta(t) $ is just the  classical equation of motion 
for the particle 
 \begin{gather} 
 \ddot{\eta} + {\omega}^2 {\eta} = 0 
 \end{gather} 
 The solution of this represents  particle position in the direction of wave propagation and  x - component of velocity 
  is solely determined by the frequency of the oscillator. 
 Particle position within WP is a complex quantity defined by y(t) is connected with WP width
 of the packet within wave packet at any time t and is given by 
 \begin{gather} 
 \cfrac{ 2 \hbar}{ m} \dot{y} + {\left(\cfrac{2 \hbar}{m} y(t)\right)}^2 + \omega^2 (t) = 0
 \end{gather} 
  Separating equation into  real and imaginary part   , we get 
  \begin{gather} 
  \cfrac{ 2 \hbar} {m} \dot{y_I} + 2 \left( \cfrac{ 2 \hbar}{m} y_I\right) \left( \cfrac{ 2 \hbar}{m} y_R\right) = 0 \\
  \cfrac{ 2 \hbar}{m} \dot{y_R} + {\left(\cfrac{ 2 \hbar}{m} y_R\right)}^2 - {(\cfrac{ 2 \hbar}{m} y_I )}^2 + \omega^2(t) = 0 
  \end{gather} 
  The real $ y_R(t)$ can be eliminated from equation ( 13 - 14 ) by solving equation (12  ) for $ y_R $ .  Inserting  new variable $ \alpha^2 $ \cite{schuch}   as
\begin{gather} 
   \cfrac{ 2 \hbar}{m} y_I = \cfrac {1}{ \alpha^2 (t) } 
  \end{gather} 
into equation for $ y_I(t) $ , one obtains a  non linear classical equation for particle position within WP as
\begin{gather} 
\ddot{\alpha} + \omega^2 \alpha = \cfrac{1} { \alpha^3} = g( \alpha(t) ) 
\end{gather} 
which is identified as   autonomous  EP equation for any positive oscillator frequency value $ \omega^2 > 0 $.  Equation (17)  is an  initial value problem with $ \alpha(0)= \alpha_0 $ . 
This equation is considered as very fundamental along with equation (12) to understand evolution of particle trajectory within quantum bound. 
Schuch et al \cite{schuch} solved  equation (17) by linearizing and defining Feynman kernel to understand classical - quantum connection .
Many DE s encountered in physics are unstable although integrable or partially integrable in some physical sense. It is then extremely
 important not to discard them and this is achieved by 
imposing some of mathematical requirements.  Kruskal \cite{kruskal} gave techniques of non linear regular - singular type analysis as sufficient conditions for Painleve property. 

 In this work ,  we developed solution as meromorphic function of the particle trajectory function in complex domain.  
  Without defining Feynman Kernel, one can solve ODE (17) which exhibits significant behavior of the particle trajectory. 
  
\section{Solution using Jacobi Elliptic Form } 

Multiplying both side of equation (17)  by $ \dot{\alpha}(t)  $ and 
solving ODE  ( with initial condition $ \alpha(0) = \alpha_0 , \dot{\alpha}(0)  = 0 $) , ODE can be reduced to first order ODE as
\begin{gather} 
 {\dot{\alpha}}^2 + {\alpha}^2 + \cfrac{1}{ {\alpha}^2} = \alpha_0^2 + \cfrac{ 1}{ {\alpha_0}^2 } 
\end{gather} 
The solution of this ODE becomes 
\begin{gather} 
t - t_0 = {\int} ^{\alpha} \cfrac{ \alpha_0 d a } { \sqrt{ ( 1-    a^2  {\alpha_0}^2 )  (  {\alpha_0}^2 - a^2  ) } }     \\
   \text{or}  \\
t - t_0 = ( {\int} ^{a= \alpha \alpha_0 } \cfrac{ d a } { \sqrt{ ( 1-  a^2  )  (  1 - \cfrac{a^2}{ {\alpha_0}^4 }  ) } } 
\end{gather}  
with $ a = a  \alpha_0 $ . 
We can write solution as
\begin{gather} 
t - t_0 = F( \phi , k) = \int  \cfrac{ d a } { \sqrt{ ( 1-  a^2  )  (  1 - k^2   a^2)}  } 
\end{gather}  
which is known as Jacobi incomplete elliptic integral  of first kind  with $ m  = k^2 $ , known as Jacobi modulus. 
In case of inverting in the integral for inverse of sine function ,  we define integrand as 
\begin{gather} 
 w  := \cfrac{ 1 } { \sqrt{ ( 1 - a^2) ( 1 - k^2  a^2) }} 
\end{gather}
 
Since ODE possess  singularity at  a =0  , we  consider  Riemann surface  $ \mathbb{P }  =  \mathbb{C} - {0} \cup {\infty} $  around zero 
( puncture corresponding to  $ \alpha( t_0) = zero $ ) for the existence of solution. 

So, right hand of integral is denoted as  incomplete integral for first kind in Legendre form given by 
   \begin{gather} 
   F( a , k) =   \int  \cfrac{ d a } { \sqrt{  ( 1-  a^2   )  (  1 - k^2   a^2 )}  } 
  \end{gather}
  and it can be expressed as Meromorphic function as 
  \begin{gather} 
   F( a , k) =   \int  \cfrac{ d a } { \sqrt{  ( 1-  a^2   )  (  1 - k^2   a^2 )}  } 
  \end{gather}
  in complex plane with $ a = \alpha_0 (  \alpha(t) - \alpha(t_0) ) $. 
   Clearly, F is branched - its value at any point depends on the path of integration due to square root singularities of integrand at  $ a = \pm 1, \pm \cfrac{1}{k} $ . Note that 
  $  F( a , 0) = sin^{-1}a $ which is inverse of entire periodic function. 
  Integrand is  represented by  Jacobi elliptic function ( given by ( ) )   on $ \mathbb{C} $ which satisfies a first order differential equation. 
With the choice for the path of integration , i.e $ F(0,k) =0 and $ F(1,k) = K , K as positive real number 
\begin{gather} 
K = {\int_0 }^1 ( 1 - a^2)^{-\cc} ( 1 - k^2 a^2 )^{-\cc} da 
\end{gather} 
is called complete elliptic integral of first kind. So, in complex plane under mapping $ a \to F( a, k) $, the image of segment of real axis $ 0 \to 1 $ is the segment of real axis $ 0 \to K $.  As a  passes the value a =1 along small semi circle  $ a = 1 - \epsilon  e^{ - i \theta } $ as $ \theta $ varies , integrand becomes pure imaginary number. 
This gives 
\begin{gather} 
F(a, k) = K + i {\int_1}^a ( 1 - a^2)^{-\cc} ( 1 - k^2 a^2 )^{-\cc} da  , 1 < a < \cfrac{1}{k} 
\end{gather}
So, the image of segment of real axis $ 1 \to \cfrac{1}{k} $ is the straight line segment $ K \to K + i K^{\prime} $ 
where 
\begin{gather} 
{\int_1}^{\cfrac{1}{k}}  ( 1 - a^2)^{-\cc} ( 1 - k^2 a^2 )^{-\cc} da 
\end{gather} 
in known as complementary integral of first kind. 
 So, we construct Riemann surface $ \mathbb{X} $  where function will behave holomorphically by 
removing two branch cuts from $ \mathbb{P} $.
such that $   \mathbb{X} := \mathbb{P} - \lbrace {1} \cup \cfrac{1}{k} \rbrace $. 
Thus , $ \mathbb{X} $ will remain  non singular upper  domain of ( a , t) on $ \mathbb{C}^2 $.  

To evaluate integral  in exact form ,  integral can be explicitly evaluated in terms of Appell function which id hypergeometric functions of two arguments. The evaluation requires 
 a generalization of binomial formula of arbitrary exponents i.e 
 \begin{gather} 
 ( 1 -a)^j = 1 - j a + \cfrac{ j(j-1) }{2 !} a^2 - \cfrac{ j(j-1) ( j -2) }{ 3!} a^3 + \cdots 
 \end{gather} 
and proceeds as follows 
\begin{gather} 
\int da ( 1- a^2)^{ -\cc} ( 1 - k^2 a^2)^{\pm \cc} \\
= \int da \cfrac{ \Gamma( 1/2 + p)  \Gamma( \pm \cc) } { \Gamma(\cc)   \Gamma( \pm \cc) }  \cfrac { k^2r}{ p! r!} a^{ 2p + 2r}  \\
= a F_1 ( \cc; \cc, \pm \cc; 3 \cc; a^2, k^2  a^2) = t - t_0 = u 
\end{gather} 

This integral can be inverted using  Lagrange inversion method 
\begin{gather} 
 a = F^{-1} ( t - t_0)  = c + {\sum_{k=1}}^{\infty}  \cfrac{ [ t - t_0 - f(c) ]^k} { k! } \lim_{ a \to c} \cfrac{ d^{k-1}}{ d a^{k-1}} \left[ \cfrac{ a -c} { f(a) - f(c)} \right]^k
\end{gather} 
The inverse of Appell function can be expressed by Bell polynomial. 
So, the inverse of incomplete elliptic integral of first kind can be expressed as 
\begin{gather} 
F^{-1} ( u , k) = u + {\sum_{n=1} }^\infty \cfrac{ u^{2 n +1}} { ( 2 n+1) n !} {\sum_{l=1}}^n b_{ n.l} B_{n, l} ( k_1, k-2, \cdots, k_{ n-l+1} ) 
\end{gather} 
 So, we can express solution as 
 \begin{gather} 
 \alpha(t) - b   = t - t_0 - \cfrac{1 +k^2 } {6} {( t - t_0)}^3 + \cfrac{ 1 + 14 k^2  + k^4} {120} ( t - t_0)^5 - \cdots 
 \end{gather}  
 The expression is normalized with respect to $ \alpha_0 $ .  This precisely matches the expansion of Jacobi elliptic sine about origin. 
 The radii of convergence for this series will be controlled by the singularities of Appell function and rough estimate can be taken as 
 $ \cfrac{1}{ k } $. 
 
 \section{Solution using Pole Expansion} 
 
 On the other hand, if we consider EP equation without order reduction, it   exhibits  singular  behavior
  as  $  t  \rightarrow  t_0 $ due to presence of  $ g( \alpha( t) )$.  So, we can consider constructing solution as  meromorphic near such point. 
  Since $ t_0 $ is a moving point whose value depends on integration constant, singularity is referred as  pole (movable singularity )  .  
  Geometrically, solution  will represent two  dimensional 
  Riemann surface  ( $ \mathbb{S}  \in \mathbb{C}^2 $).  Since Riemann surface can represent  graph of a well defined polynomial function in the domain , 
  we can use this surface  to construct 
  any complex analytic function and projection can be used to $ ( \alpha(t), t  ) $ plane in the neighborhood of  
  $  ( \alpha(t_0), t_0) $ as non singular domain.  
  A Riemann surface  $  \mathbb{S} \in \mathbb{C}^2 $ is non singular , if each point can be used to identify
   a neighborhood of   $  ( \alpha(t_0), t_0) $  on $ \mathbb{S} $ 
  homomorphically 
  with   two dimensional disk $ t - \alpha (t) $ plane above  $ t_0 , \alpha(t_0) $  point.  
Using following two theorems , we construct Laurent series for the function on Riemann disk. 
\begin{thm} 
  If $ f: {\Omega_p}^{\star} $  holomorphic,  then for  $ D_R(p) $  ( centered around p)  satisfying  $ \overline{ D_R(p)}  \subset \Omega $ , 
  f(z) can be represented by Laurent series 
  \begin{gather} 
  f(z) := { \sum_{- \infty} }^{\infty} a_j ( z - p)^j \\ \text{or} 
    f(z) := {\sum_{j=1}}^{\infty} \cfrac{ b_j} { ( z- a )^j } + {\sum_{j=o}}^{\infty} c_j ( z-a)^j 
  \end{gather}    
  \end{thm}      
and                                          
  \begin{thm} 
  For above Laurent series, there exist  $ R_1 $ , $ R_2$ such that 
  series converges on the disk for all $ z \in \mathbb{C} $ satisfying 
 \begin{gather} 
 \cfrac{1}{ R_1} \mid z - a \mid R_2 
\end{gather} 
Moreover, convergence is uniform/absolute in the region $  r_1 \le \mid z - a \mid \le r_2 $ for any $ r_1 $ and $ r_2 $ satisfying
$ \cfrac{1}{R_1} < r_1 < r_2 < R_2 $ . As a consequence, the limiting function is holomorphic in the annulus.
\begin{gather} 
R_1 = ( \lim_{ j \to \infty}  \mid b_j \mid^{ 1/j} )^{(-1) } \\
R_2 =  ( \lim_{ j \to \infty}  \mid c_j \mid^{ 1/j} )^{(-1) }
\end{gather}
\end{thm} 
So, we can represent general solution of $ \alpha(t) $ as Laurent series 
\begin{gather} 
f : \alpha(t)  = {\sum_{ j=0}}^{\infty}   a_{j +p}  {( t - t_0) }^{j +p} \\   \text{with} 
f(p) : = \infty 
\end{gather} 
in extended complex plane  as  $ f: \Omega \to \tilde{\mathbb{C} } = \mathbb{C} \cup \infty $ .
Here Riemann surface is centered around $ \alpha(t_0) :  \mid  \alpha - \alpha( t_0) \mid  < \epsilon $.  
 Since order of pole p  depends on $ t_0 $ value ( or integration constant ), therefore  p is  essential simple pole. 
   The principal part of the series  can be used to calculate  on  substitution  into ODE yields 
 \begin{gather} 
  [  p(p-1) {a_0 }  \tau^{p-2} +  p(p+1)  {a_1}  \tau^{p-1}  + \cdots ]  + {\omega}^2 [ {a_0} {\tau}^p  +
 {a_1}  \tau^{p+1} + \cdots ] - [  \cfrac{1}{{a_0}^{3} } \tau^{-3p}  + \cfrac{1}{ {(a_1) }^3}  { \tau}^{ -3(p+1)} + \cdots ] =0 
 \end{gather} 
 Having only poles as movable singularities is a special case of Painleve  property that is required by an ODE to have general analytic solution around 
 such pole which is determined through following  theorem 
\begin{thm} 
A point $ z = z_0 \in \mathbb{C} $ of function  $ y(z) : \mathbb{C} \rightarrow \mathbb{C} $ is a pole of order p if 
\begin{gather} 
y(z) = { \sum_{i= -p} }^{\infty} \alpha_i {( z - z_0) }^i , p \in \mathbb{N} 
\end{gather} 
and A point $ z = z_0 $ of function y(z) is an algebraic branch point /singularity if 
\begin{gather} 
y(z) = {\sum_{i= -p}}^\infty \alpha_i { ( z - z_0)}^{i/n} , p, n \in \mathbb{N} 
\end{gather}    
if n=1, then it is a pole.          
\end{thm}     
To calculate order of pole p, we express ODE  in equivalent form
  \begin{gather} 
  E [( \alpha(t) ] =   \ddot{\alpha} + {\omega}^2 {\alpha} -  \alpha^{-3}  = 0 \\ \text{or} 
  E = { \sum_{ j=0} }^{\infty} E_j \tau^{ j +q} 
  \end{gather} 
Analyzing and matching dominant power term  for both equation ( ) and ( )  on basis of Puisseux  diagram  yields ( p, q) = ( -1, -3)  
with condition that p must take  negative integer value
corresponding to negative integer value of q. 
So, series representation is following 
\begin{gather} 
\alpha(t)  = {\sum_{ j=0}}^{\infty}   a_j  {( t - t_0) }^j +  \cfrac{a_{-1 } }  {  t - t_0 }
\end{gather} 
Near pole, solution behavior indicates that it captures generic singular solution of the ODE ( since solution has two arbitrary constants $ a_{-1} , t_0 $ ). 
  \begin{gather} 
 \alpha( t) =   \cfrac{ a_{-1}  } { t - t_0} 
 \end{gather}
In general, solution behavior can be expressed as 
\begin{gather} 
\alpha(t) = {\sum_{j=0}}^{\infty} a_{j -1} ( t - t_0)^{j-1}   \text{or} \\
\alpha( t) =  a_{-1}  ( t - t_0)^{ -1} + a_0 + a_1 ( t -t_0) + a_2 {( t - t_0)}^2 + \cdots 
\end{gather} 
with $ a_{-1}  $ as fundamental coefficient of the series.  The value of principal coefficient is basically residue of the function near pole. 

To evaluate  all  other  coefficients, one has to substitute  ( 25 ) into ( 16) and perform leading term balance  to yield 
 \begin{gather} 
 a_{-1} = c* i \\
 a_0 = 0 \\
 a_1 = - \cfrac{2}{3} \omega^2 a_{-1} \\
 a_2 = - \cfrac{ a_{-1} a_1} { 6 a_1 + 4 a_{-1} a_1 + 2 \omega^2 a_{-1} } \\
 a_3= \cfrac{ 1 - 2 {a_{-1}}^2 {a_2}^2 + 2 ( 1 + 3 \omega^2){ a_{-1}}^2 {a_1}^2 } { 2 {a_{-1}}^2 a_1 + 4 {a_{-1}}^2 + 4 \omega^2 {a_{-1}}^3 } \\ 
 \end{gather} 
So, meromorphic Laurent series is following 
 \begin{gather} 
  \alpha(t) =  a_{-1}  \cfrac{ 1}  { t - t_0}  -   \cfrac{2}{3} \omega^2 a_{-1}  ( t - t_0)  -  \cfrac{ a_{-1} a_1} { 6 a_1 + 4 a_{-1} a_1 
  + 2 \omega^2 a_{-1} } {( t - t_0)}^2 \\  \nonumber
  +  \cfrac{ 1 - 2 {a_{-1}}^2 {a_2}^2 + 2 ( 1 + 3 \omega^2){ a_{-1}}^2 {a_1}^2 } { 2 {a_{-1}}^2 a_1 + 4 {a_{-1}}^2 + 4 \omega^2 {a_{-1}}^3 } {( t - t_0) }^3  + \cdots
    \end{gather} 
   It is to note that fundamental coefficient is  imaginary quantity for any arbitrary real number c and higher order coefficients appear 
   in terms of fundamental coefficient. This  indicates particle trajectory is confined in complex plane of  
 Riemann punctured annulus surface  of radius $   \vert t - t_0 \vert  < \delta $  corresponding to $ \vert \alpha(t) - \alpha( t_0) \vert < \epsilon   \in \Omega $. 
 In this case, pole has simple point structure. 
 Kruskal \cite{kruskal} pointed out that even if the only possible formal solution of such ODE is Laurent series,
  the poles indicated by these series may accumulate elsewhere to give rise to a worse branched singularity. ODE can turn unstable again and again in time space in fact. 
We assume here that oscillator frequency   ( $ \omega $ ) 
does not take  any singular value.  This  indicates singularity arises in ODE (16 )  only due to  $  g( \alpha( t))  $.
Clearly, $ \alpha(t) $ has branched movable essential singularity .   
Most of the non linear  ODEs possess pole like singularity ( movable because it depends on integration constant).

\section{Calculation of exact meromorphic solution} 
Both solution ( ) and ( ) is holomorphic on Riemann surface. 
 Demina et al \cite{demina} showed that if stable  Laurent series exist for a holomorphic function f for all $ z \in \mathbb{C} $ : 
  then f  can be explicitly represented by 
  simply periodic  function  with period T ( function having pole of order p inside a strip of periods built on Riemann surface ) by 
\begin{gather} 
f(z) := \cfrac{ \pi} { T} \{  {\sum_{k=1}}^p \cfrac { {(-1)}^{k -1} c_{ -k} d^{ k -1} }{ ( k -1) ! d {(z- z_0)}^{ k -1} } \}  \ cot \left( \cfrac{ \pi z} { T} \right ) + h_0 
\end{gather} 
  Proof of their theorem is based on  Mittag - Leffer expansion for meromorphic function  and is given in \cite{demina}  
which yields f for  our case ( with p=1) 
 \begin{gather}
f:=  \alpha(t) = \alpha( t) =  - a_{-1}  \sqrt{L}   \cot \left( \sqrt{L}  (  t - t_0)  \right) + h_0 
 \end{gather} 
 is doubly periodic meromorphic function for any constant term   $ h_0 $.  L is related to time period  T of  meromorphic function by     $ L = \cfrac{ \pi^2}{T^2} $. 
The function can be expressed in terms of Laurent series   as
 \begin{gather} 
\alpha( t) =  -  \cfrac{a_{-1}  }{t - t_0}  + \cfrac{1}{3}  a_{-1}  L(  t - t_0) + \cfrac{1}{45} a_{-1}  L^2 {( t - t_0)} ^3 + \cfrac{2}{945} a_{-1}  L^3 {( t - t_0)} ^5 + \cdots  higher orders   + h_0                                                                                                                                                                                                                                                                                                                                                                                                                                                                                                                                                                                                                                                                                                                                                                                                                                                                                                                                                                                                                                                                                                                                                                                                                                                                                                                                                                                                                                                                                                                                                                                                                                                                                                                                                                                                                                                                                                                                                                                                                                                                                                                                                                                                                                                                                                                                                                                                                                                                                                                                                                                                                                                                                                                                                                                                                                                                                                   
\end{gather} 
Comparing coefficients of equation ( 55  ) with (52)  , we can obtain value of T in terms of fundamental coefficients 
\begin{gather} 
 T ={ \pi } \sqrt[4]{\cfrac{a_{-1}} {45}} 
 \cfrac{1}{\sqrt[4]{\cfrac{ 1 - 2 {a_{-1}}^2 {a_2}^2 + 2 ( 1 + 3 \omega^2){ a_{-1}}^2 {a_1}^2 } { 2 {a_{-1}}^2 a_1 + 4 {a_{-1}}^2 + 4 \omega^2 {a_{-1}}^3 }} }
  \end{gather}
  Since value of  T depends on $a_{-1} $, so can take real /complex . 
So, we can write general meromorphic solution of the particle trajectory  near singularity as
\begin{gather} 
 \alpha( t) =  a_{-1}   \cfrac{\pi}{T}  cot \left( \cfrac{ \pi ( t - t_0)}{T}   \right)  + h_0  - \alpha_0 
\end{gather} 
as cotangent plane with time period T given by ( 56 ). 

\section{Conclusion}
  Both solution ( 35) and (58) remain  analytic in the punctured Riemann surface . That indicates that particle trajectory is only confined in punctured Riemann surface as its quantum bound. 
  The existence of meromorphic solution near movable singularity of the ODE confines particle trajectory in  of punctured complex Riemann surface. 
This is  behavior of a classical particle in quantum domain.
   Since $ \alpha(t) = \sqrt{ \cfrac{ m}{ 2 \hbar y_I(t) }} $  is connected to position uncertainty of the particle within wave packet,  $ \alpha( t \rightarrow t_0)    \ne  0 ) $ condition is necessary in order to have single valued function. 
Since $ \alpha(t) $  can not take zero value as point of singularity , it means  position uncertainty of  particle  that remains finite   at any time t. Or, in other words,  particle will remain confined on Riemann surface within gaussian wave packet determined by frequency of the oscillator. 
Solution for  $ \alpha(t) $ is also branched near singularity. Solution using both approach yields periodic function in complex plane which matches with wave function solution from Schr$\ddot{o}$edinger equation. 
    
  \section{Acknowledgement}
  Author Ruma Dutta thanks Prof Stan for giving computational facility at Department of Ohio State University.


\begin{thebibliography}{99} 
 
 \bibitem{abra} 
 Abramowitz M. \& Stegun I.A ., {\it Handbook of Mathematical Functions with Formulas , Graphs and Mathematical Tables}, {\bf Chapter 17, National Bureau of Standards, Washington}, (1964). \\
 
 
\bibitem{andri}
Andripoulos K. \& leach P.G.L., {\it The economy of complete symmetry groups for linear higher- dimensional systems}, {\bf \ J. \ Non linear \ Math. \ Physics 9, Second Supplement }, 10 - 23, (2002). \\

\bibitem{bul} 
Bulirsch R., {\it Numerical computation of elliptic integrals and elliptic functions II}, {\bf \ Numer. \ Math. 7}, 353 - 354, ( 1965b). \\

\bibitem{byrd} 
Byrd P.F \& Friedman M.D., {\bf  Handbook on Elliptic Integrals for Engineers and Physicists, 2nd ed., Springer , Berlin} , (1971). \\

\bibitem{cruz} 
Cruz H., Schuch D. , Castanos O. \& Ortiz O. R., {\it Time Evolution of quantum systems via a complex non linear Riccati equation.I. Conservative systems with time - independent Hamiltonian} , { \bf 
arXiv:1505.02687v1 [ quant-ph] , May }, 22 pages, ( 2015). \\

\bibitem{erm} 
Ermenko A., {\it Meromorphic traveling wave solutions of the Kuramato-Sivahinsky equation}. {\bf \ Math. \ Phys. \ Anal. \ Geom. 2(3)}, 278 - 286, (2006). \\

\bibitem{demina} 
Demina M.V \& Kudryyashov N. A., {\it From Laurent Series to Exact Meromorphic Solutions: the Kawahara Equation}, {\bf https://arxiv.org/abs/1112.5266}, (2011). \\
2011. \\

\bibitem{eli}
Eliezer C.J \& Gray A.A.; {\it  A note on the time dependent harmonic oscillator}, {\bf  \ SIAM. \ J. \ Appl.  \ Math}, {\bf 30}, 463 - 468, 1976. \\

\bibitem{erma}
Ermakov V.; { \bf Second - order differential equations, Conditions of complete integrability}, {\it Universitetskie Izvestiye}, {\bf 1880(9)}, 1-15, 1880. \\

\bibitem{fukushima} 
Fukushima T., {\it Fast computation of incomplete elliptic integral of first kind by half argument transformation}, {\bf \ Numer. \ Math. 116}, 687 - 719, ( 2010). \\


\bibitem{hoyer} 
Hoyer P., {\it Uber die Integration eines Differential gleichungss systems von der Form durch elliptische Funktionen}, {\bf Dissertation Kongeigl , Friedrich 
- Wilhelms Univ, Berlin}, 1- 36, (1879). \\


\bibitem{Joos} 
Joos E., {\it Decoherence and the Appearance of a Classical World in Quantum Theory}, {\bf Second Ed. Berlin}, 41 - 88, (2203). \\

\bibitem{kleber} 
Kleber M., {\it Exact solutions for time - dependent phenomena in quantum mechanics}, {\bf Physics Reports 236(6)}, 331- 393, (1994). \\


\bibitem{kovalevski} 
Kovalevski S.V., {\it Sur le probleme de la rotation d'un corps solide autour d'un point fixe}, {\bf \ Acta \ Math 12}, 177 - 232, (1889). \\

\bibitem{kruskal} 
Kruskal M.D., Joshi N. \& Halburd R., {\it Analytic and Asymptotic Methods for Non Linear Singularity Analysis: A Review and Extensions of Tests for the
Painleve Property}, {\bf \ Integrability of Non Linear Systems 495}, 171 - 205, (2007). \\

\bibitem{labahn} 
Labahn G. \& Humphries T.D., {\it  Symbolic Integration of Jacobi Elliptic Functions in Maple}, {\bf https://cs.uwaterloo.ca/~glabahn/Papers/jacobi-elliptic.pdf}. \\
 
 \bibitem{leach} 
 Leach P.G.L \& Andriopoulos K., {\it The Ermakov equation , A commentary}, { \bf \ Appl. \ Discrete \ Math. (2)}, 148 - 157, (2008). \\
  
 \bibitem{lewis}
Lewis H.R Jr. \& Risenfeld W.B.; {\bf An exact quantum theory of the time-dependent harmonic oscillator and of a charged particle in a time dependent electromagnetic field},
{\it  \ J. Math. \ Phys.}, {\bf 10}, 1458-1473, 1968. \\

\bibitem{morris} 

Morris A.H. Jr., {\it NSWC Library of Mathematical Subroutines}, {\bf   Tech. Prep. NSWCDD/TR-92/425, NAval Surface warfare Center, Dahlgren}, 107 -110, (1993). \\


\bibitem{pickering} 
Pickering A., {\it The singular manifold method revisited}, {\bf \ J. \ Math. \ Phys. 37(4)}, 1894 - 1927, ( 1996). \\

\bibitem{pinney}
Pinney E.; {\it The Non Linear Differential Equation $ y^" + p(x) y + c y^{-3} = 0 $ }, {\bf  Proc. Amer. Math. Soc.}, {\bf 1}, 681, ( 1950) . \\

\bibitem{siu} 
Siu Y.T., { \it Elliptic Functions ( Approach of Abel and Jacobi ) } ,{ \bf  http://people.math.harvard.edu/siu/}, 1- 77,  (2020). \\

\bibitem{schuch} 
Schuch D. , {\it On the relation between the Wigner function and an exact dynamical invariant} , {\bf \ Phys.  \ Lett.  A 338}, 225 - 231, (2005). \\


\bibitem{pr1} 
Press W.H., Teukolsky S.A., Vetterling W.T. \& Flannery B.P., {\it Numerical Recipes}, {\bf The Art of Scientific Computing, Cambridge Univ Press }, (1986). \\


\bibitem{pr2} 
Press W.H., Teukolsky S.A., Vetterling W.T. \& Flannery B.P., {\it Numerical Recipes}, {\bf The Art of Scientific Computing, Cambridge Univ Press , 3rd ed},(2007). \\



\bibitem{ziglin}
Ziglin S.; {\bf branching of solutions and nonexistence of first integrals in hamiltonian mechanics I \& II }, 
{\it  Functional  Analysis \& Its   Applications }, {\bf 17}, 6 - 17, (1983).  \\

\bibitem{ziglin2} 
Ziglin S., { it Self - intersection of the complex separatices and the nonexistence of the integrals in the Hamiltonian systems with one -and 
half degrees of freedom}, {\bf \ J. \ Appl. \ Math 45}, 411- 413, (1982). \\

\bibitem{yoshida} 
Yoshida H., {\it Necessary condition for the existence of algebraic first integrals I: Kovalevski's exponents}, {\bf \ Celestial \ Mechanics 31}, 363 - 379, ( 1983). \\


\end{thebibliography}
\end{document}